\documentclass{jetpl}
\twocolumn

\lat

\usepackage{hyperref}

\title{Dark matter mass fraction in axion stars}
\rtitle{Mass fraction of axion stars}
\sodtitle{Mass fraction of axion stars}

\author{A.S. Dmitriev$^{\; a}$, D.G. Levkov$^{\; a}$, A.G. Panin$^{\; a,c}$, I.I.\,  Tkachev$^{\; a,d}$ \thanks{e-mail: tkachev@inr.ru}    }
\rauthor{Dmitriev A., Levkov D., Panin A., Tkachev I.}
\sodauthor{Dmitriev, Levkov, Panin, Tkachev}

\address{$^a$Institute for Nuclear Research of the Russian Academy of Sciences,
  Moscow 117312, Russia\\
$^c$Moscow Institute of Physics and Technology, Dolgoprudny 141700, Russia\\
$^d$Physics Department and Laboratory of Cosmology and Elementary Particle Physics, Novosibirsk State University, Novosibirsk 630090, Russia}

\abstract{
QCD axion dark matter is fragmented into small mass miniclusters of high density.  With the natural parameters of miniclusters, the time for Bose-Einstein condensation is less than the age of the Universe, which leads to the formation of Bose stars inside these structures. Recently, the problem of the subsequent growth of Bose stars was solved analytically~\cite{Dmitriev:2023ipv}. We use this solution to determine present day mass fraction of dark matter in Bose stars and their mass function.
}

\dates{XXX}{XXX}

\begin{document}
\maketitle

\paragraph{1. Introduction.}
\label{sec:intro}

If the dark matter particles are light bosons, such as QCD axions or their ``fuzzy'' axion-like cousins~\cite{RingwaldPDG}, then compact macroscopic objects, Bose stars~\cite{Ruffini:1969qy, Tkachev:1986tr} (for a recent review see~\cite{Visinelli:2021uve}),  are expected to form as a result of  Bose-Einstein condensation. Remarkably, large couplings are not needed, and the condensation time can be less than the age of the Universe with typical parameters of the substructures where they nucleate, simply as a result of universal gravitational interaction~\cite{Levkov:2018kau}. Therefore, Bose-stars may be abundant in mainstream dark matter models.  The phenomenological implications of their existence, such as micro-lensing~\cite{Kolb:1995bu, Fairbairn:2017sil, Ellis:2022grh}  or radioemission~\cite{Tkachev:1986tr, Levkov:2020txo, Witte:2022cjj, Kouvaris:2022guf, Escudero:2023vgv, Kyriazis:2023ehv}, which potentially ends up in the FRB-like phenomena~\cite{Tkachev:2014dpa, Iwazaki:2014wka, Raby:2016deh, Buckley:2020fmh}, and, consequently, the associated prospects~\cite{Eby:2017xaw} for indirect dark matter searches  strongly depend on the present day Bose star mass function.

We define mass function in a similar way as stellar mass function is defined. Namely, it gives the fraction of dark matter in Bose stars of a given mass. A simple way to estimate it is to use the phenomenological core-halo relation~\cite{Schive:2014hza} if the halo mass function is known. That had been done for fuzzy axion dark matter $m_a \sim 10^{-11}$\;eV in Ref.~\cite{Du:2023jxh}. At the higher levels of presision we need to know how the mass of a particular star grows from its nucleation to the present, depending on the environment and initial conditions. This later problem had been solved analytically recently~\cite{Dmitriev:2023ipv}.

In this Letter we calculate Bose star mass function  for the QCD axion dark matter  model. Bose stars in this model are increasingly referred to as axion stars. They nucleate  inside the smallest gravitationally bound virialized cosmological structures, which are miniclusters of axions~\cite{Kolb:1993hw, Vaquero:2018tib, Buschmann:2019icd, Eggemeier:2019khm, Ellis:2022grh,Ellis:2020gtq,Pierobon:2023ozb}. The binding energy is tiny, so the constituent particles are deeply non-relativistic.

\paragraph{2. Nucleation and growth of a Bose star.}
\label{sec:nucl}
  
At small particle masses $m$ the occupation numbers of dark matter axions are large, so their clump can be
described by a random classical field $\psi(t,\, \boldsymbol{x})$
evolving in its own gravitational potential~$U(t,\,  \boldsymbol{x})$,
\begin{align}
  \label{eq:GP2}
  &i \partial_t \psi = -\Delta \psi/2m  + mU \psi \,, \\
  \notag
   &\Delta U = 4\pi G\,  (m|\psi|^2 - \bar{\rho}) \,.
\end{align}
In the Poisson equation $\bar{\rho} \equiv M/L^3$ is the mean mass density and $M$ is the total mass of a minicluster. For simplicity, here we  approximated the structure with the periodic box of size $L$.  The starting point of its evolution is a virial equilibrium, i.e.\ a Gaussian-distributed field with Fourier image $|\psi_{\boldsymbol{p}}|^2 \propto \mathrm{e}^{-\boldsymbol{p}^2/p_0^2}$ and random phases~$\arg\psi_{\boldsymbol{p}}$, where ~$\omega_0\equiv p_0^2/2m$ is the typical particle energy. 

This ensemble  thermalizes into a Bose-Einstein condensate with nucleation of a Bose-star. On average, in a virialized halo, this occurs in kinetic time~\cite{Levkov:2018kau}  
\begin{equation}
 \label{eq:tgr}
t_{gr} = b\, \frac{4\sqrt{2}}{27\pi \Lambda}\, \frac{R}{v} \left( mvR \right)^3 ,
\end{equation} 
where $\Lambda = \ln(mvR)$ is the the Coulomb logarithm,  and $M,R,v$ are virialized halo values for the mass, radius and velocity dispersion respectively, $ v^2 = \langle {\bf v}^2\rangle \approx   GM/R$. With this choice of parameters  ${b \approx 0.03}$. To avoid confusion, we note that in Ref.~\cite{Levkov:2018kau} the value ${b \approx 0.7}$ is given, but there the central density and typical velocity $p_0/m$ were chosen to parametrize the miniclusters.
 \begin{figure}
   \centerline{\includegraphics[width=\columnwidth]{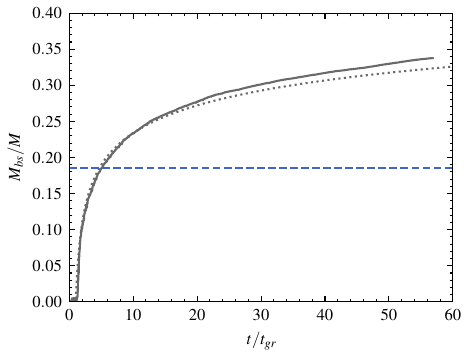}}
  \caption{{\bf Figure 1.} Bose star mass $M_{bs}(t)$ for $\epsilon\approx 0.186$.  Solid line corresponds to the  result of numerical simulation of the system (\ref{eq:GP2}), while dotted -- analytical solution Eq.~(\ref{eq:3}).  Horizontal dashed line shows core-halo level.}
  \label{fig:Mbs_t}
\end{figure}

The subsequent evolution of the ensemble turns out to be self-similar, which makes it possible to find analytical solutions~\cite{Dmitriev:2023ipv}. The results were confirmed by numerical simulations in the periodic box and isolated miniclusters. Remarkably, the derived evolution equation is determined by the generic scalings of the gravitational collision integral, while the solution  for the Bose star mass involves only conservation laws. Then, the dependence of a Bose star mass,  $x_{bs} (\tau) \equiv M_{bs}/M$, with time is given by~\cite{Dmitriev:2023ipv}
\begin{equation}
  \label{eq:3}
  (1 + x_{bs}^3/\epsilon^2)^3 (1 - x_{bs})^{-5} \approx  t/t_{gr}\ \,.
\end{equation}
where 
\begin{equation}
  \label{eq:defeps}
\epsilon^2 \equiv E/\gamma M^3  = v^2/2\gamma M^2  \, ,
\end{equation}
and is fixed by the initial minicluster mass $M$ and 
energy $E$, while $\gamma = c_\gamma m^2 G^2 $ contains a model dependence with $c_\gamma \approx 0.0542$ being a numerical constant defined in Ref.~\cite{Dmitriev:2021utv}.  When using the virial velocity for $v$, Eqs.~(\ref{eq:3}, \ref{eq:defeps}) agree well with the results of our numerical simulation of the growth of a Bose star in a virialized halo.

Goodness of this solution is illustrated in Fig.~\ref{fig:Mbs_t} for a particular value of $\epsilon \approx  0.186$ in the periodic box simulation, see also~\cite{Dmitriev:2023ipv}. We see that the theory~(\ref{eq:3}) (red dotted curve) closely reproduces simulation results for $M_{bs}$ (solid).

The $\epsilon$ parameter is  primary  and determines the growth of a Bose star in dimensionless time $t/t_{gr}$. At the same time, the widely used core-halo relation~\cite{Schive:2014hza} boils down to $M_{bs}/M = \epsilon$, see~\cite{Dmitriev:2023ipv}. This corresponds to the level where the fast initial linear growth of a Bose star mass is followed by a slower increase, see Fig.~\ref{fig:Mbs_t}.  Many stars in a limited duration simulation will likely be close to this level.  But this is only a rough approximation. 

In what follows, it is convenient to factorize $\epsilon$ in $t_{gr}$.  First, using the mass-radius relation for a virialized self-gravitating cloud, $v^2 \approx GM/R$, and definition of $\epsilon$, we find $mvR = 1/( \epsilon\, \sqrt{2c_\gamma} )$. This tells us that at $\epsilon \gtrsim 1$ Bose stars nucleate in just a free fall time, see Eq.~(\ref{eq:tgr}), and gives the Coulomb logarithm $\Lambda = - \ln(\epsilon \sqrt{2c_\gamma} )$.  Overall, the right hand side of Eq.~(\ref{eq:3}) can be written in terms of $\epsilon$ as
\begin{equation}
  \label{eq:tutgr}
\frac{t}{t_{gr}} = \frac{t}{t_{u}} \left(\frac{\epsilon}{\epsilon_0}\right)^6\Lambda \,,
\end{equation}
where $t_{u}\approx 1.38\cdot 10^{10}$\,yr is the age of the Universe and
\begin{equation}
  \label{eq:eps0}
\epsilon_0^6 \equiv \frac{b}{27\sqrt{2}\, \pi c_\gamma^3\, t_u G^2 m^3 M^2}  \, .
\end{equation}
Therefore, the second parameter which governs  Bose stars growth rate is related to $t_0^{-1} \equiv G^2 m^3 M^2$. 

\paragraph{3. Minicluster parametrisation.}

We measure axion and minicluster masses in terms of benchmark parameters $m_{10} \equiv m/(10\, \mu\rm eV)$ and $M_{13} \equiv M/(10^{-13} M_\odot)$ correspondingly.  Minicluster mass and radius are related as~\cite{Kolb:1993hw}
\begin{equation}
\label{eq:MR}
  R = \frac{1.6\cdot 10^7 M_{13}^{1/3} }{\Phi(1+\Phi)^{1/3}} \; \rm km \, ,
\end{equation}
where $\Phi$ is a particular minicluster overdensity over mean axion density $\rho_a$, namely $\Phi \equiv (\rho_{MC} - \rho_a)/\rho_a$. Prior to the time when a minicluster becomes gravitationally self-bound and separates from the cosmological expansion, $\Phi$ is constant. Later on it determines a minicluster density in terms of $\rho_a$ at the matter-radiation equality~\cite{Kolb:1993hw}. This epoch appears because $\Phi \sim 1$ typically, and can be much larger than~1, i.e. miniclusters become gravitationally self-bound at around matter-radiation equality. Together with the definition  $\mu \equiv m_{10}M_{13}^{2/3}$ this gives minicluster parameters which govern axion star nucleation and growth rates
\begin{equation}
\label{eq:epsilon}
  \epsilon = 0.0389\, \mu^{-1} \, \Phi^{1/2}(1+\Phi)^{1/6} \, ,
\end{equation}
\begin{equation}
\label{eq:epsilon0}
\epsilon_0 \equiv 0.0215\, \mu^{-1/2}\, .  
\end{equation}

 \begin{figure}
   \centerline{\includegraphics[width=\columnwidth]{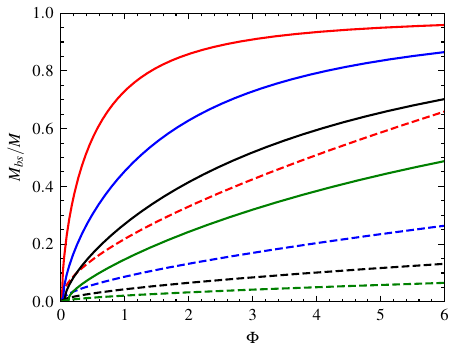}}
   \caption{{\bf Figure 2.} The ratio of axion star mass  at present $M_{bs}(t_u)$ to the initial mass of the minicluster is shown by the solid lines  for several values of $\mu$ (from top to bottom $\mu = 0.2,0.5,1,2$) as a function of $\Phi$.  Dashed lines of the same color show result based on the core-halo relation. }
  \label{fig:Mbs_traj}
\end{figure}

Using now Eq.~(\ref{eq:3}) we find the fraction of the minicluster mass that by now has condensed into an axion star. It is shown by the solid lines in Fig.~\ref{fig:Mbs_traj} as a function of $\Phi$ for several values of $\mu$. We see that the larger is $\Phi$, the larger is condensed fraction, and light miniclusters (small $\mu$) condense almost entirely regardless of $\Phi$. Dashed lines of the same color show rough approximation based on the core-halo relation. The difference is significant.

\paragraph{4. Axion stars mass function.} 

To calculate present day Bose star mass function on the basis of Eq.~\ref{eq:3} we need minicluster distribution over $\Phi$ and $\mu$, or, at fixed axion mass, over   $\Phi$ and $M$, which we will call $P(\Phi, M)$.  It is defined as
\begin{equation}
\label{eq:PMC}
P(\Phi, M) \equiv \frac{1}{F_0}\frac{d^2\, F}{d\Phi  \,  d\log M},
\end{equation}
where $F$ is a fraction of dark matter in miniclusters. In what follows the normalization factor $F_0$ is chosen in such a way that $\int  P(\Phi, M) {d\Phi  \,  d\log M} = 1$.

In fact, $F$  is time dependent. There are several reasons for that. 
First, at the QCD epoch, when miniclusters are born, axion mass is temperature dependent and  changes with the Universe expansion, while miniclusters with the most interesting large $\Phi$ values are in fact long-living oscillons that  decay late.   Finial $\Phi$ depends not only on the magnitude of the  initial ``perturbation'', but also on the moment when an  oscillon decays and axion mass freezes out. 

Second, already in the linear stage with respect to the  axion filed, miniclusters gravitationally merge forming larger structures~\cite{Eggemeier:2019khm, Ellis:2022grh}. Here two outcomes are possible. A) Two miniclusters can really merge loosing identities and forming more massive object, like two spiral galaxies merge into a larger elliptical one. B) Alternatively, they can merge forming cluster of miniclusters, like galaxies form clusters of galaxies.  For our purposes of studying the formation of axion stars, these class B encounters, in which identities are not lost, should not be counted in changing the distribution of $P(\Phi, M)$ at smaller scales.

Third, minuclusters are destroyed in collisions with usual stars in the Galaxy. However, only small fraction of them is destroyed during the age of the Galaxy~\cite{Tinyakov:2015cgg, 2017JETP..125..434D, Kavanagh:2020gcy, Dandoy:2022prp, Shen:2022ltx}.  We can  neglect this effect. Destruction of formed axion stars is even more difficult. On the other hand we cannot take $P(\Phi, M)$ at a very late time since we need to trace the process of Bose condensation from the time when it becomes efficient.

Considering what was said, it looks reasonable to take $P(\Phi, M)$ at the moment of matter-radiation equality and map it through Eq.~\ref{eq:3} into axion star distribution. Doing that we will loose an effect of close encounters of the type A, but  account of those should be based on dedicated large scale numerical simulations and we feel it should be the subject of a separate study. 

In the early study~\cite{Kolb:1995bu} the reduced distribution $\tilde P(\Phi) = \int P(\Phi,M)\, d\log M$ was calculated, and not far from the epoch of minicluster formation, i.e. at a fourfold decrease in temperature. Nevertheless, an agreement with recent studies in the region of  $\tilde P(\Phi) \sim 1$ is reasonable. In Ref.~\cite{Vaquero:2018tib} minicluster distribution had been evolved up to the factor of 5 decrease in temperature, but again, only $\tilde P(\Phi) $ is presented. Late time distribution had be found in~\cite{Eggemeier:2019khm} but only   $\tilde P(M)$ is given on the contrary.  The most suitable for our  needs are the results of~\cite{Buschmann:2019icd} were $P(\Phi, M)$ was calculated at the moment of matter-radiation equality. This function is presented in \cite{Buschmann:2019icd} as a two-dimensional color plot, which we digitized. In the same work, the axion mass yielding the correct dark matter abundance was calculated as $m_a = 25 \pm 11 \;\mu{\rm eV}$, in agreement with~\cite{Klaer:2017ond}.

\begin{figure}
   \centerline{\includegraphics[width=\columnwidth]{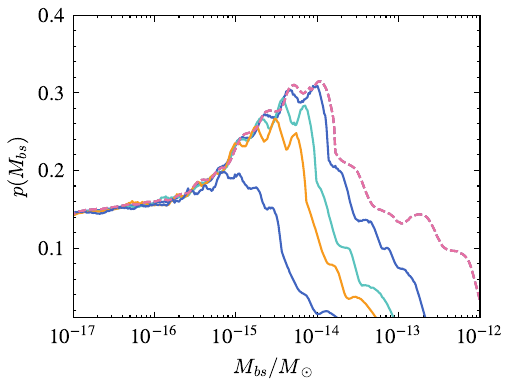}}
   \caption{{\bf Figure 3.} The fraction of the total mass of miniclusters contained in axion stars at the present stage of the evolution of the Universe is presented as a function of the mass of these stars. Solid lines from top to bottom  correspond to  $m_a =10, 25, 40, 100\; \mu{\rm eV}$. The dashed line shows the minicluster mass function used.
 }
  \label{fig:Mbs_dist}
\end{figure}

We define present day QCD axion dark matter mass function in axion stars  similarly to Eq.~(\ref{eq:PMC})
\begin{equation}
\label{eq:Pbs}
p(M_{bs}) \equiv \frac{d\, F_{bs}}{ d\log M_{bs}}.
\end{equation}
Its calculation is straightforward.  We evolve each point $(\Phi, M)$ up to the present moment according to equation (\ref{eq:3}) (as  in Fig. \ref{fig:Mbs_traj}) and add the condensed fraction $(M_{bs}/M)\,P(\Phi, M)$ to the bin corresponding to the found value of $M_{bs}$.
The result for four values of the axion mass is shown in Fig.~\ref{fig:Mbs_dist}  by solid lines. The dashed line corresponds to $P(\Phi, M)$ from Ref.~\cite{Buschmann:2019icd} integrated over $d\Phi$. It is normalized to unity. We see that at $m_a =10$\;$\mu$eV miniclusters with $M  \lesssim 10^{-13}M_\odot$ are almost completely condensed by now, while for  $M \lesssim 10^{-15}M_\odot$ this happens regardless of the axion mass value. The total condensed mass fractions of all miniclusters for the quintet of axion masses as in Fig.~\ref{fig:Mbs_dist} are equal to $(0.86, 0.74, 0.67,0.51)$, respectively.

\paragraph{5. Discussion.}
\label{sec:discussion}

In this Letter we have calculated present day QCD axion star masses and their distribution employing recently found~\cite{Dmitriev:2023ipv} analytical solution for the Bose star mass growth, Eq.~(\ref{eq:3}). Figure~\ref{fig:Mbs_traj} shows condensed mass fraction of a virialized minicluster depending upon its parameters: initial mass and overdensity. This result does not depend on the  uncertain details of the minicluster distribution and can itself be useful for a variety of applications.

The mass fraction of dark matter in axion stars can be obtained by multiplying the distributions shown in Fig. \ref{fig:Mbs_dist} by $F_0$.  According to Refs.~\cite{Kolb:1995bu, Vaquero:2018tib, Buschmann:2019icd, Eggemeier:2019khm}  $F_0$ is in the range 0.5 -- 0.75.  However, since in our approach $p(M_{bs})$ is tied to  $P(\Phi, M)$  calculated at the epoch of matter-radiation equality, it therefore does not take into account  the type A  mergers of miniclusters, coalescence of previously formed axion stars,  and minicluster destruction in collisions with usual stars.  The fraction of disrupted miniclusters in the latter process is not large, however. Only a few percent of miniclusters with $\Phi \sim 1$ crossing the solar orbit are destroyed during the existence of the Galaxy, the rest remain intact regardless of the minicluster mass~\cite{Tinyakov:2015cgg}.

The survival of miniclusters through mergers with larger halos in the early epoch was studied in Ref.~\cite{Ellis:2022grh}. Miniclusters with higher mass lose a smaller mass fraction, but even for miniclusters with low mass, the core of the density profile is preserved, and the tail dissipates, cf.~\cite{Berezinsky:2007qu}. In addition, their fate has been traced in N-body cosmological simulations, while wave equations~(\ref{eq:GP2})  are required to study the survival of axion stars formed at early stages in low-mass miniclusters.

Also, our Eq.~(\ref{eq:3}) does not account for the secondary infall of diffuse axions on the initial minicluster. Accumulated particles will be less bound, in a fragile halo~\cite{Sikivie:1996nn}, and will be the first to dissipate in encounters. So, both unaccounted effects, infall and disruption, partially compensate each other.

Modeling the formation and evolution of miniclusters is very complex and must involve very different spatial and temporal scales at different cosmological epochs to resolve axion strings, domain walls, and oscillons. This is reflected in the uncertainties in $P(\Phi, M)$.  In particular, its values in the high $\Phi$ high $M$ tail are unknown. Formation of overcritical and collapsing axion stars~\cite{Levkov:2016rkk} in this corner may lead to spectacular phenomena~\cite{Tkachev:2014dpa, Levkov:2020txo} even if  $P(\Phi, M)$ is tiny and does not contribute to Fig.~\ref{fig:Mbs_dist}, Also, recently, the uncertainties on the derived axion mass had been increased, $m_a = {\cal O} (10 - 100)\;  \mu{\rm eV}$, reflecting difficulties in estimation of the axion strings contribution, see~\cite{Pierobon:2023ozb} and references therein.

However, despite all the uncertainties discussed, we believe that our results provide useful first  insight on the QCD axion star mass function.

\paragraph{Acknowledgements.}

We thank K. Dolgikh for discussions. This work was supported by the grant RSF 22-12-00215-П. 



\bibliographystyle{ieeetr} 

\bibliography{refs} 

\end{document}